\newcommand{\be}{\begin{equation}}
\newcommand{\ee}{\end{equation}}
\newcommand{\bea}{\vspace{0.25cm}\begin{eqnarray}}
\newcommand{\eea}{\end{eqnarray}}
\documentclass{ws-ijqi}
\usepackage{graphics}
\usepackage{epsfig}
\usepackage{graphicx}

\catchline{}{}{}{}{}

\title{Quantum bit error rate in modeled atmospheres.}

\author{ N. Antonietti \footnote{nicolo.antonietti@polito.it}, M. Mondin }
\address{Dipartimento di Elettronica, Politecnico di Torino, corso Duca degli Abruzzi, 24, 10129 Torino, Italy}

\author{ F. Daneshgaran}
\address{California State University Los Angeles, ECE Department, 5151 state university drive, Los Angeles CA 90032, USA}

\author{G. Giovanelli, I. Kostadinov, B. Lunelli }
\address{CNR, ISAC, via Gobetti, 101, 40129 Bologna, Italy}

\author{ G. Brida, M. Genovese, M. Gramegna}
\address{INRIM, Strada delle Cacce, 91, 10135 Torino, Italy}

\begin{document}

\maketitle
\begin{history}
\end{history}

\begin{abstract}
The quantum bit error rate is a key quantity in quantum
communications. If the quantum channel is the atmosphere, the
information is usually encoded in the polarization of a photon. A
link budget is required, which takes into account the depolarization
of the photon after its interaction with the atmosphere as well as
absorption, scattering and atmospheric emissions. An experimental
setup for the reproduction of a simple model of the atmosphere is
used to evaluate the quantum bit error rate in a BB84 protocol and
the results are presented. This result represents a first step
toward the realization of an optical bench experiment where atmospheric effects
are simulated and controlled for reproducing the effects on a
quantum channel in different meteorological situations.
\end{abstract}

\keywords{QBER; modeled atmosphere.}

\section{Introduction}
Recent experiments have proven the feasibility of quantum communication
in free space. In \cite{schmitt-manderbach:010504}, an experimental
implementation of quantum key distribution over a 144 km free-space
link is reported. In \cite{1367-2630-10-3-033038}, the first
experimental study of the conditions for the implementation of the
single-photon exchange
between a satellite and an Earth-based station is accounted for.\\
The former experiment was located at a mean altitude of more than
2400 m, along an horizontal path over the sea and exploited a
quantum communication protocol relying on decoy-states. Under
good atmospheric conditions (where the authors do not specify the atmospheric parameters),
a 10 dB attenuation was
assumed to be due to atmospheric losses; with the further assumption
that the eavesdropper could not exploit multiphoton pulses, a 28
bit/s bit rate and a 6.77\% QBER were obtained. The latter
experiment was carried out along a vertical downward path (from a
satellite to the Earth); a 81\% atmospheric transmission was
estimated in accordance with the modeled atmosphere losses
(according to \cite{antonietti-2007-17}, they match the losses in a
clear atmosphere, at the experimented wavelength of 532 nm). A 157
dB total attenuation was observed and 4.6 counts per
second were estimated.\\
The first remark we want to make on both experiments is the low bit
rate: it is too low compared to the present request which may be as
high as many Mb/s; thus neither of these two experimental results
are useful, so far, to exchange a quantum key under desired
operating conditions. Second, the quantum communication security
decreases as the bit rate gets closer to the dark counts rate. For
instance, in a BB84 protocol, the highest losses for a secure
quantum transmission were estimated to be 40 dB\cite{rarity:240},
which is far less than the losses obtained in
\cite{1367-2630-10-3-033038}. Thus, a lot of work must be directed
towards an improvement of the quantum key bit rate and
of the error rate of the communication link.\\
To this end, enhancements can be made in the source and receiving
devices, or studies can be carried out on the quantum communication
channel and the latter option is investigated in this paper.\\
Furthermore, the realization of an operative Earth -- space quantum
channel will require the possibility using the system under
different meteorological situations (e.g., in presence of haze or
translucent clouds, such as cirri clouds) and when the satellites is
at different angles from the horizon. Thus, the operativity of the
system must be estimated under all these conditions.

Our purpose is to investigate and record the losses under different
atmospheric conditions. Those will contribute, together with the
other sources, to the final losses and to the final bit rate. Thus,
before the actual communication takes place, one is able to know whether the quantum
key exchange is feasible (by estimating the bit rate) and secure (by
estimating the total losses), depending on the atmospheric
conditions.\\
A full-scale experiment has two main drawbacks:

\begin{enumerate}
    \item it is expensive,
    \item it can not be created on demand.
\end{enumerate}

Those disadvantages can be overcome by experimentation in
laboratory. In this paper, we report on our first approach to
model different atmospheres in laboratory and to estimate the QBER
for a simple BB84 quantum communication protocol. We present
preliminary results showing the feasibility of this approach,
motivating the interest toward the realization of more complex
systems where different atmospheric situations are really simulated
and controlled.

\section{Measurement setup}
\begin{figure}[h]
\begin{flushleft}
  \includegraphics[width=15cm]{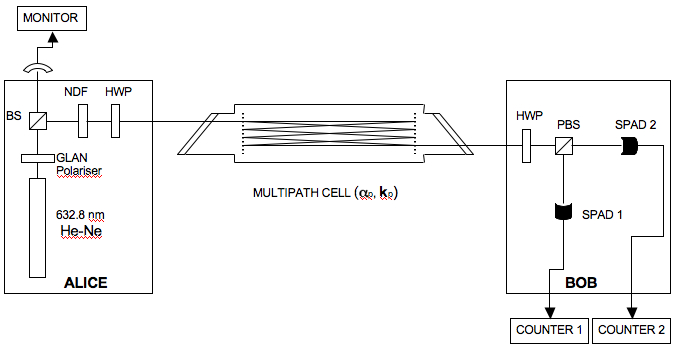}
  \caption{Schematic of a BB84 quantum communication protocol. Alice
  is the source of photons. The light of a laser is plane-polarized after the
  GLAN polarizer and its power is then monitored by measuring the fraction of
  light split in a beam splitter (BS). The power of the light is attenuated by means
  of neutral density filters (NDF) and the plane of polarization is chosen by means
  of a half-wave plate (HWP). The photons in the multipath cell extend their path based
  on the number of reflections inside the cell; a gas with an extinction coefficient $k$\ is inserted
  into the multipath cell to simulate the atmosphere. Finally the photons reach the receiver Bob. A Half-wave
  plate (HWP) and a polarizing beam splitter (PBS) allow Bob to choose the bases which the photons are measured in.
  Two single photon avalanche diodes (SPAD) and their counters measure the number of photons in the
  two different states of plane polarizations (either horizontal or vertical).}
  \label{BB84_schem}
\end{flushleft}
\end{figure}

A BB84 quantum communication protocol \cite{bb84-orig,gisin-2001} experiment
was mounted in laboratory, as depicted in figure (\ref{BB84_schem}).

Alice is the source of single photons, Bob is the receiver and the
multipath cell is the quantum channel.\\
The source of light is a 11.3 mW output power helium-neon laser,
with a 500:1 plane polarization ratio and a $\lambda = 632.8$ nm
wavelength. The laser beam is attenuated by means of neutral density
filters down to the photon counting regime. Ideally, single photons
are supposed to be used; however, the available ones (PDC heralded
photons, quantum dots, color centers in diamond) are not ideal and
they can be approximated by low--average-photon--number coherent
states (weak laser beams), with the only side effect of a smaller
security threshold.

A polarizer guarantees a better polarization ratio of at least
1000:1.
Finally a half-wave plate is used to rotate the plane of polarization of the photons.\\
A multi-path cell was first described by John U. White in
1942\cite{White:42}. The bulk of a White cell consists of three
spherical concave mirrors having the same radius of curvature and
positioned to form an optical cavity. Depending on the mutual
position of the mirrors, the light is subject to many reflections
inside the cavity, hence extending its total path.\\
At the exit of the cell, the light is directed into a half-wave
plate, which can twist the plane of polarization again. A polarizing
beam splitter (hence PBS) splits the beam into two different optical channels;
the input light, according to the relative polarization, and the
exit beams are measured by two identical SPADs, namely the PDM 5CTC model by
MPD-Micro Photon Devices with quantum efficiency of 38\% @633 nm
and 78 ns dead-time.

\section{The experiment}

\begin{figure}[h]
\begin{flushleft}
  \includegraphics[width=15cm]{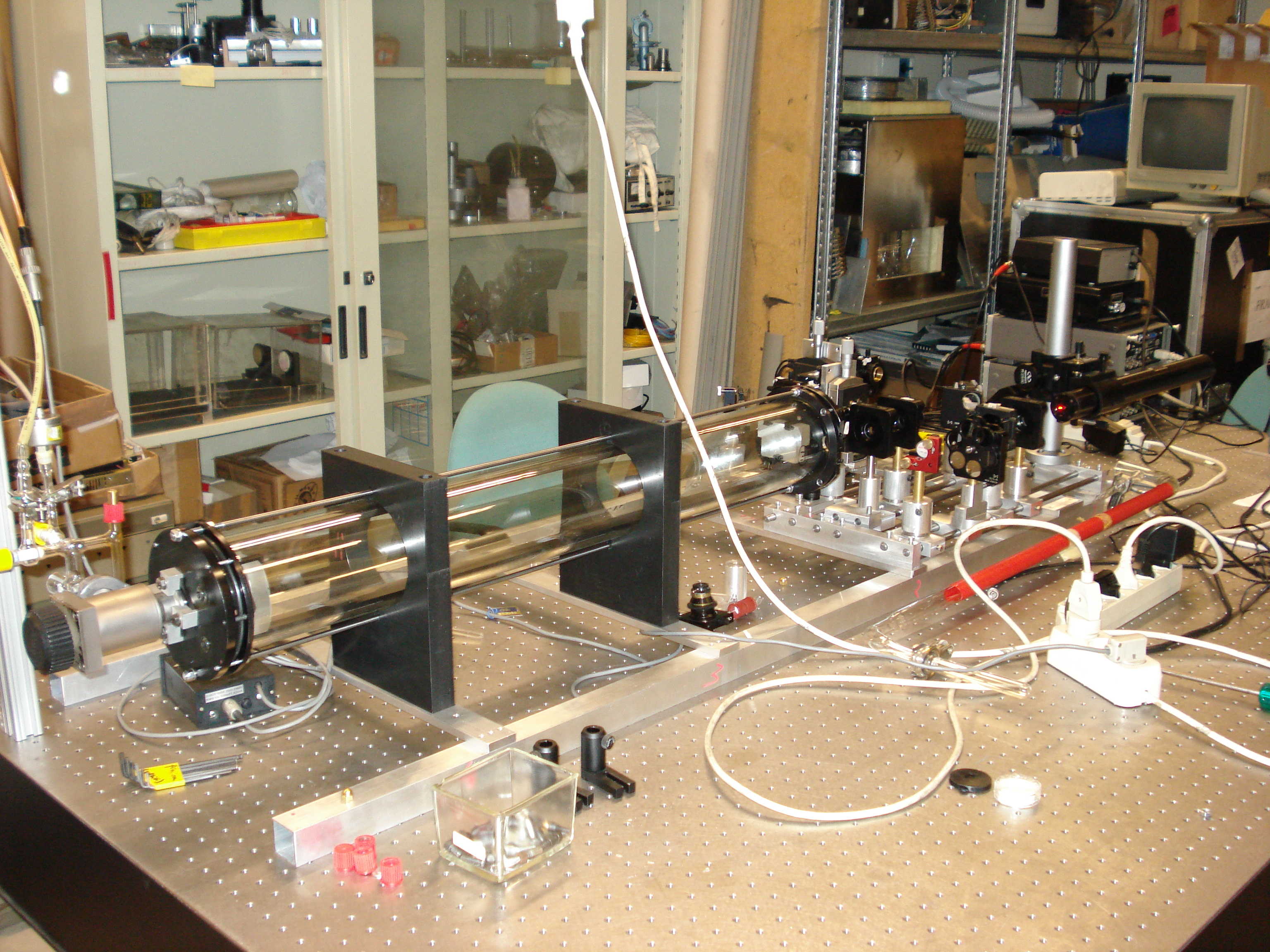}
  \caption{Picture of the experimental setup assembled in laboratory to
  test a BB84 protocol through the atmosphere. In the foreground, the transparent
  multipath cell is seen and in the background the source and receiver are placed.}
  \label{apparatus}
\end{flushleft}
\end{figure}

\begin{figure}[h]
\begin{flushleft}
  \includegraphics[width=15cm]{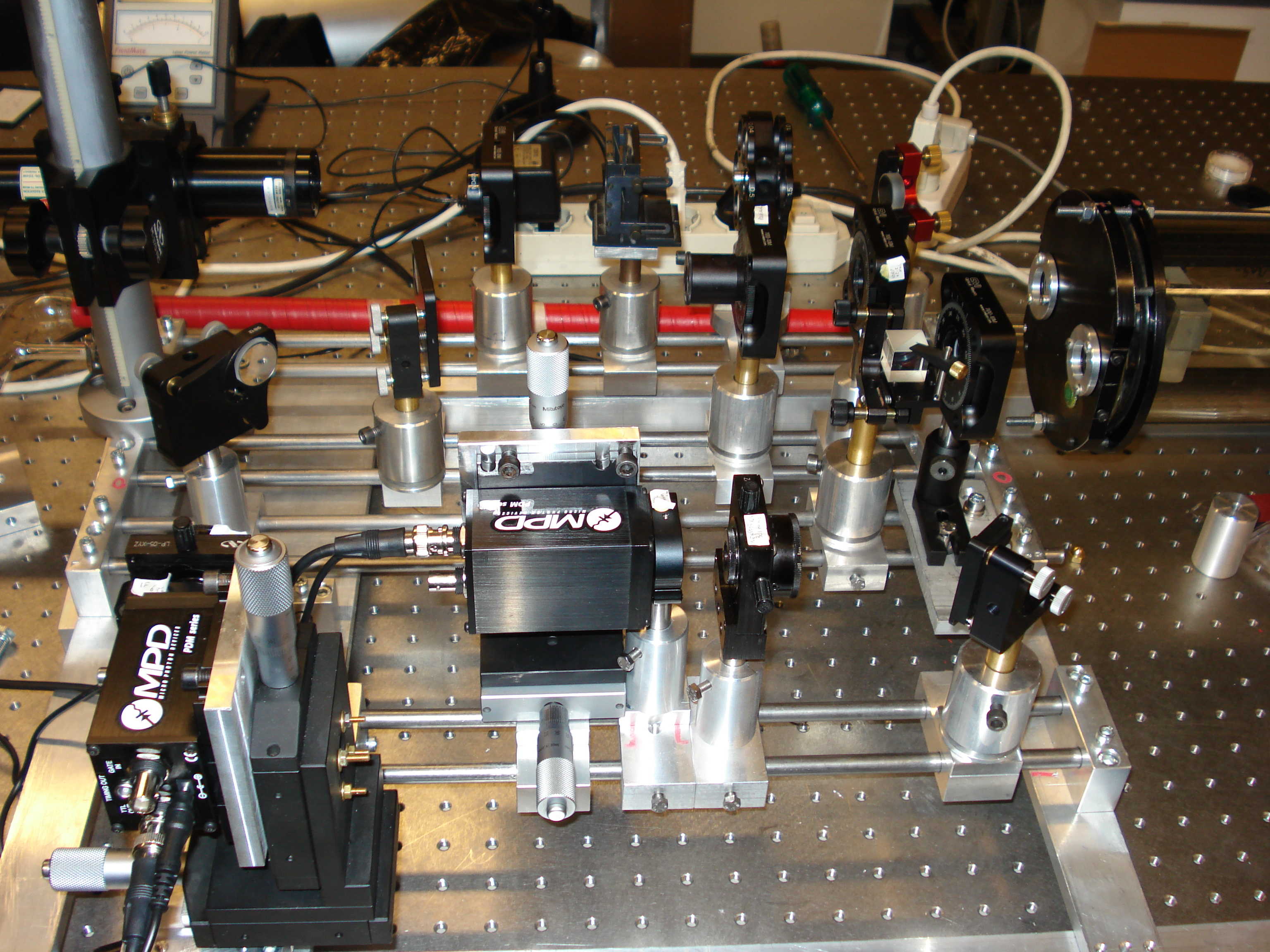}
  \caption{Detail of the experimental setup: the source is placed far from the camera,
  while the receiver setup can be seen closer.}
  \label{detail}
\end{flushleft}
\end{figure}

Both Alice and Bob dispose of half wave plates and polarizers for
setting the transmission and measurement bases (for a presentation
of our set-up see fig. \ref{apparatus} and fig. \ref{detail}).

Alice can simulate 4 states of polarization of light by means of
four rotations of her half-wave plate:

\begin{enumerate}
    \item vertical: the half wave plate is in its vertical position,
    \item horizontal: the half-wave plate is rotated by $45^o$
    clockwise,
    \item left diagonal: the half-wave plate is rotated by $22.5^o$
    counter clockwise,
    \item right diagonal: the half-wave plate is rotated by $22.5^o$
     clockwise.
\end{enumerate}

Bob needs two bases: one vertical and horizontal (herein VH), the
other, left diagonal and right diagonal (herein LR). The PBS
before the single photon counters is a VH basis; in order to have a
LR basis, the half-wave plate in front of the PBS is rotated by $22.5^o$ clockwise.

In  BB84 protocol, the polarization of the transmitted photon and
the basis of the measured photon are randomly chosen. Since our
purpose to investigate the effects of the atmosphere on a
communication link, we chose to set the halfwave plates in each one
of the eight possible positions and to measure
the number of photons in each of the two directions.\\
In a first configuration, the multipath cell was set to vacuum state
and the inside mirrors were tilted so that, after many reflections,
the path was 22.4 m long \footnote{It is worth mentioning that much
longer paths can be obtained with other cells.}. Considering the
time discrimination window of the detector equal to their dead-time
(78 ns) we arranged the attenuation of the neutral density filters
placed in front of the laser so that, for a vertical polarization
and a VH basis, the count rate on the single photon receivers were
of the order of $10^5$ photons/s. Under this condition the average
number of photons within the time discrimination window of a
detector
is $\mu \approx 7.8 \cdot 10^{-4}$\ photons. \\
As we stated before, we intend to demonstrate that in a lab it is
possible to simulate the effect of the atmosphere on a quantum
communication protocol; the main effect of the atmosphere is the
absorption of the signal which lowers the useful photon counts
making it more and more equal to the dark counts and lowering the
reliability of the communication link. Thus, as a second step,
bromine was injected into the cell and the measurements were
repeated. Bromine is chosen because it has a high molar absorptivity
at $\lambda = 632.8$\ nm; that will reduce the useful signal making
it comparable to the dark counts. In such a case, the error rates
would be more significant and the effect of the atmosphere can be
shown.

\section{Results}

The QBER is defined as the ratio of the amount of photons detected
with the wrong polarization to the total amount of detected photons.
It derives both from optical imperfections and from the effects
of the communication channel. \\

When the cell is placed in vacuum state (the pressure is less than 1
hPa), the mean measured QBER was 0.86\%. This is only due to
imperfections of the optical instruments. The polarization ratio of
the polarizer is 1000:1, so we can consider our light in a state of
plane polarization for our purposes.\\
Afterwards, 30 ml of bromine was injected into the cell (the bromine
molar absorptivity is $\epsilon = 1.3\ cm^{-1}(mol/ L)^{-1} $\
\cite{PasschierA.A._j100863a025}). In the vacuum, it evaporates,
filling the whole cell, until a pressure of 26 hPa is reached.
The measurements are then repeated.\\
We first noticed that, only 1\% of the previous amount of photons
are transmitted through the bromine dioxide and counted by the SPADs. \\
In this condition the QBER was 7.68\%. A value approaching the
security limit of BB84 which is 11 \%, a result showing the feasibility of
a simulation of atmospheric effects in lab.\\

Let us now consider the extinction coefficient $k$.
 If $P_0$\ is the power of the beam at the receiver when there is
vacuum in the cell and $P$\ is the power of the beam at the receiver
with the bromine inside the cell, we
measured $\frac{P}{P_0}= 0.01$.\\
 We can then evaluate the length $L$\ traveled by
the light in different atmospheres (assuming a parallel plane
atmosphere \cite{1986afgl.rept.....A}), corresponding to the values
of $k$\ we set in order to obtain the same attenuation,
$\frac{P}{P_0}= e^{-kL}$.  This allows to estimate what QBER is to
be expected in different atmospheres, before the quantum
transmission. We have matched the path in an atmosphere to the path
in our cell full of bromine, by means of the extinction coefficient.
For instance, we can consider an horizontal transmission at ground
level under the following atmospheric conditions:

\begin{itemize}
    \item in a summer atmosphere, with urban aerosols and 5 km
    visibility, $k = 2.62 \cdot 10^{-1} \mbox{ km}^{-1}$\ and $L= 17.6 \mbox{ km}$;
    \item in a summer atmosphere, with urban aerosols and 13 km
    visibility, $k = 9.01 \cdot 10^{-2} \mbox{ km}^{-1}$\ and $L= 51.1  \mbox{ km}$;
    \item in a summer atmosphere, with rural aerosols and 5 km
    visibility, $k = 4.60 \cdot 10^{-2} \mbox{ km}^{-1}$\ and $L= 100.07 \mbox{ km}$;
    \item in a summer atmosphere, with rural aerosols and 13 km
    visibility, $k = 1.58 \cdot 10^{-2} \mbox{ km}^{-1}$\ and $L= 290.9 \mbox{ km}$;
    \item in a winter atmosphere, with urban aerosols and 5 km
    visibility, $k = 2.54 \cdot 10^{-1} \mbox{ km}^{-1}$\ and $L= 18.1 \mbox{ km}$;
    \item in a winter atmosphere, with urban aerosols and 13 km
    visibility, $k = 8.38 \cdot 10^{-2} \mbox{ km}^{-1}$\ and $L= 52.7 \mbox{ km}$;
    \item in a winter atmosphere, with rural aerosols and 5 km
    visibility, $k = 4.51 \cdot 10^{-2} \mbox{ km}^{-1}$\ and $L= 102.1 \mbox{ km}$;
    \item in a winter atmosphere, with rural aerosols and 13 km
    visibility, $k = 1.55 \cdot 10^{-2} \mbox{ km}^{-1}$\ and $L= 270.0 \mbox{ km}$.
\end{itemize}
The description of the different kinds of aerosols used can be found
in \cite{1990apuv.agar.....S} and the values of the relative
extinction coefficient can be obtained by simulation with the
MODTRAN program\cite{1993SPIE.1968..514A}.

\section{Conclusions}
We reported on our first approach to model a simple atmosphere in
laboratory. The reproduced atmosphere was used to test a BB84
quantum communication protocol in free space. As a preliminary
result, we measured the effect
of that atmosphere in terms of the absorption and QBER.\\

Future steps will provide a better description of real atmospheres
and the evaluation of their effects on quantum communication in
situations representing various meteorological conditions.

\subsection{ Acknowledgments}
This work has been supported by EU project QuCandela, by Compagnia
di San Paolo and by Regione Piemonte (E14).

\bibliographystyle{prsty}
\bibliography{bibli}

\begin{thebibliography}{10}

\bibitem{schmitt-manderbach:010504}
T. Schmitt-Manderbach {\it et~al.}, Physical Review Letters {\bf 98},  010504
  (2007).

\bibitem{1367-2630-10-3-033038}
P. Villoresi {\it et~al.}, New Journal of Physics {\bf 10},  033038 (12pp)
  (2008).

\bibitem{antonietti-2007-17}
N. Antonietti {\it et~al.}, LASER PHYSICS {\bf 17},  1389  (2007).

\bibitem{rarity:240}
J.~G. Rarity {\it et~al.},  in {\em Quantum communications in space}, edited by
  R.~E. Meyers and Y. Shih (SPIE, PO Box 10 Bellingham WA 98227-0010 USA,
  2004), No.~1, pp.\ 240--251.

\bibitem{bb84-orig}
C.~H. Bennett and G. Brassard,  (IEEE, 10662 Los Vaqueros Circle P.O. Box 3014
  Los Alamitos CA 90720-1264 USA, 1984).

\bibitem{gisin-2001}
N. Gisin, G. Ribordy, W. Tittel, and H. Zbinden, Quantum Cryptography, 2001.

\bibitem{White:42}
J.~U. White, J. Opt. Soc. Am. {\bf 32},  285  (1942).

\bibitem{PasschierA.A._j100863a025}
A.~A. Passchier, J.~D. Christian, and N.~W. Gregory, Journal of Physical
  Chemistry {\bf 71},  937  (1967).

\bibitem{1986afgl.rept.....A}
G.~P. {Anderson} {\it et~al.}, Technical report (unpublished).

\bibitem{1990apuv.agar.....S}
 in {\em AGARD, Atmospheric Propagation in the UV, Visible, IR, and MM-Wave
  Region and Related Systems Aspects 14 p (SEE N90-21907 15-32)}, edited by
  E.~P. {Shettle} (AGARD, 7115 Standard Drive Hanover , Maryland 21076-1320
  USA, 1990).

\bibitem{1993SPIE.1968..514A}
G.~P. {Anderson} {\it et~al.},  in {\em Proc. SPIE Vol. 1968, p. 514-525,
  Atmospheric Propagation and Remote Sensing II, Anton Kohnle; Walter B.
  Miller; Eds.}, Vol.~1968 of {\em Presented at the Society of Photo-Optical
  Instrumentation Engineers (SPIE) Conference}, edited by A. {Kohnle} and W.~B.
  {Miller} (SPIE, PO Box 10 Bellingham WA 98227-0010 USA, 1993), pp.\ 514--525.

\end{thebibliography}

\end{document}